# Verifying Component and Connector Models against Crosscutting Structural Views

Shahar Maoz
School of Computer Science
Tel Aviv University, Israel

Jan Oliver Ringert, Bernhard Rumpe
Software Engineering
RWTH Aachen University, Germany

## ABSTRACT

The structure of component and connector (C&C) models, which are used in many application domains of software engineering, consists of components at different containment levels, their typed input and output ports, and the connectors between them. C&C views, presented in [24], can be used to specify structural properties of C&C models in an expressive and intuitive way.

In this work we address the verification of a C&C model against a C&C view and present efficient (polynomial) algorithms to decide satisfaction. A unique feature of our work, not present in existing approaches to checking structural properties of C&C models, is the generation of witnesses for satisfaction/non-satisfaction and of short natural-language texts, which serve to explain and formally justify the verification results and point the engineer to its causes.

A prototype tool and an evaluation over four example systems with multiple views, performance and scalability experiments, as well as a user study of the usefulness of the witnesses for engineers, demonstrate the contribution of our work to the state-of-the-art in component and connector modeling and analysis.

## Categories and Subject Descriptors

D.2.1 [**Software Engineering**]: Requirements/Specifications; D.2.2 [**Software Engineering**]: Design Tools and Techniques

## 1. INTRODUCTION

The structure of component and connector (C&C) models consists of components at different containment levels, their typed input and output ports, and the connectors between them. C&C models are used in many application domains of software engineering, from cyber-physical and embedded systems to web services to enterprise applications, as they offer a physically distributed computation model as well as a logically distributed development process.

In recent work [24] we have presented *component and connector views*, as a new means to specify structural properties of component and connector models in an expressive and intuitive way. C&C views take advantage of novel abstraction mechanisms for hierarchy and connectivity, not present in comparable languages. These mechanisms allow different stakeholders to create views that express their partial knowledge about the structure of the system at hand, corresponding to different use cases, functions, or concerns.

C&C views provide means to abstract away direct hierarchy, direct connectivity, port names and types. Specifically, a C&C view may not contain all components and connectors (and typically indeed contains only a small subset of the set of all components and connectors of the system, related only to a specific use case or set of functions or features). It may contain (abstract) connectors between components at different, non-consecutive containment levels, and it may provide incomplete typing information, i.e., components' ports may be un-typed. While the standard structural abstraction and specification mechanisms supported by existing languages and tools for C&C models rely on the traditional, implementation-oriented hierarchical decomposition of systems to sub-systems, C&C views allow one to specify properties that crosscut the boundaries of sub-systems. Most importantly, this makes them especially suitable to reflect the partial knowledge available to different stakeholders involved in a system's design.

We consider two usage scenarios. First, where a set of C&C views serves as a specification for a C&C model. In this setup, different teams develop separate views, expressing constraints derived from their (partial) knowledge of the system under development. An architect is given this set of C&C views, describing mandatory, alternative, and negative structural properties, and is responsible for building a C&C model that satisfies them. Second, where a set of C&C views is created to document an existing C&C model. In this setup, the views highlight design decisions and document how specific concerns are addressed using potentially crosscutting solutions in the model.

Unlike the views, a C&C model is complete: it includes all components and connectors, with all ports names and types. It is ready for implementation, e.g., for direct code

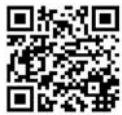

95

generation. Thus, given a C&C model, one is interested in verifying whether it satisfies each of the C&C views in its specification or documentation. However, as also demonstrated in our evaluation (see Sect. 5), manually verifying C&C views satisfaction is a difficult and error prone task.

In this work we focus on the verification of C&C models against C&C views, and present three contributions:

- First, we define and implement an efficient (polynomial) algorithm for the structural verification of a C&C model against a C&C view.
- Second, we extend the verification algorithm to not only decide satisfaction, but also, importantly, to generate small model witnesses and short natural language texts that formally justify and explain the verification results to the engineer.
- Finally, we report on an evaluation of our work over several example C&C model systems, taken from different sources and of different domains, and several C&C views specifications consisting of many views, both in terms of the performance and scalability of our algorithms, and in terms of its usefulness to engineers.

In our previous work [24], we have introduced C&C views and discussed the synthesis problem: given a C&C views specification, consisting of mandatory, alternative, and negative views, construct a concrete satisfying C&C model, if one exists. Synthesis is powerful, but it suffers from scalability limitations. The present paper complements this previous work by focusing on the dual problem of verification.

As a concrete language for C&C models we use MontiArc [2,17], a textual ADL developed using MontiCore [21], with support for direct Java code generation (including interfaces, factories, etc.). Its expressive power is comparable to that of other ADLs, e.g., MathWorks Simulink [25], AADL [13], and UML Component diagrams. The C&C views are defined as an extension to general C&C models. The concrete syntax used in our implementation is an extension of MontiArc.

Some previous work deals with verification of structural properties of component and connector structures, mainly in the context of architectures (e.g., [6,8,14]). None, however, consider witnesses that explain the verification results and none have reported on performance and scalability. We discuss these and other related work in Sect. 6.

Sect. 2 gives a semi-formal overview of C&C views and verification using examples. Sect. 3 provides formal definitions and Sect. 4 describes our verification algorithm and its output. Implementation and evaluation are presented in Sect. 5. Sect. 6 discusses related work and Sect. 7 concludes.

## 2. OVERVIEW

We use an example of a pump station architecture given with the AutoFOCUS3 tool [1,19]. We show the complete pump station C&C model and four C&C views, which we have created based on documented use cases related to the pump station system. We use these views here to demonstrate the features of the language and the relationship between a C&C model and a view.

The example is intentionally small, to support readability. Additional example systems we have used in our evaluation are briefly described in Sect. 5. We describe the example semi-formally. Formal definitions are given in Sect. 3.

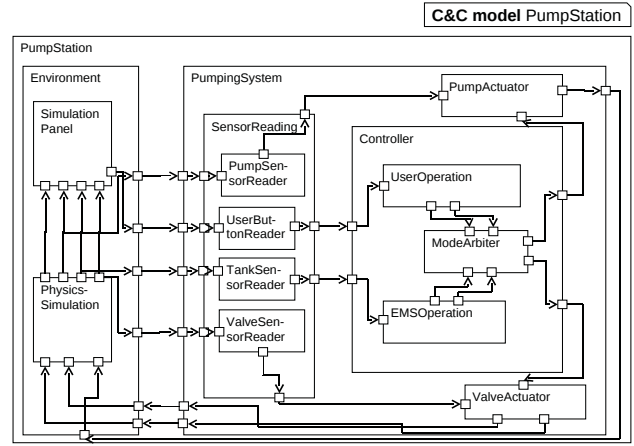

**Figure 1:** The C&C model of the pump station. Note that many tools show models like the above only one level and one sub-system at a time. Here we show the C&C model with its complete depth in one figure, in order to give a comprehensive perspective, to save space, and to contrast it with the partiality of the views as shown in other figures. Also note that to avoid clutter we omit port names and types from the figure. However, as this is a C&C model (and not a view), all ports have names and types. For example, the type of the upper left incoming port of the component `ModeArbiter` (with a connector coming from the component `UserOperation`) is `Boolean` and its name is `userPumpState`.

### 2.1 A C&C Model and Four C&C Views

Fig. 1 shows the complete C&C model of the pump station. It consists of 16 components, a containment hierarchy of 4 levels, and 49 connectors. The complete C&C model in textual, MontiArc format is available with supporting materials from [3].

Fig. 2 (left) shows a C&C view named `ASPumpingSystem`. This view focuses only on the connections between sensors and actuators in the system. As `ASPumpingSystem` is a view, it does not contain all components and connectors. While the components shown inside the `PumpingSystem` component must actually be inside it, they may be nested within some of its subcomponents (not shown in this view). Finally, each of the sensors shown must be connected to the corresponding actuator in the model, but the connection between them is not necessarily direct. It is easy to see that the C&C model satisfies the view (denoted `PumpStation` $\models$ `ASPumpingSystem`): all components mentioned in the view are present in the C&C model, the C&C model respects the containment relation specified by the view, and all connectors in the view have corresponding (chains of) connectors in the C&C model.

Fig. 2 (right) shows another C&C view, named `UserButton`. The view focuses on the components and the connectors that participate in a specific use case, namely, when the user presses the button. Again, not all components are shown, but only the ones participating in this use case. Note that the containment hierarchy is not shown, and the connectors are abstract, i.e., they specify connections, but not necessarily direct ones. It is easy to see that the C&C model satisfies this view: all the components mentioned in the view exist in the C&C model, and all abstract connectors in the view have corresponding chains of concrete

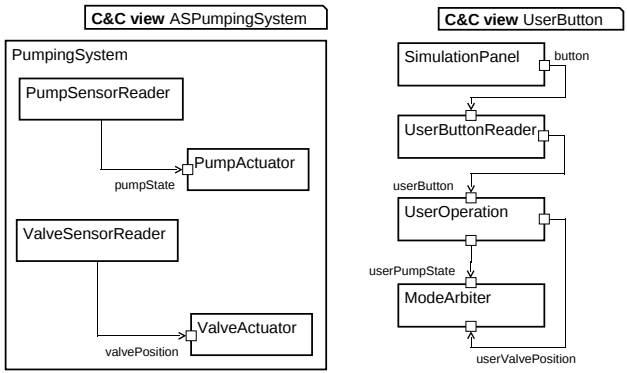

**Figure 2:** Two C&C views: `ASPumpingSystem` and `UserButton`. Note that as these are views, they allow one not to fully specify ports, port types, and port names. For example, the abstract connector going out from `PumpSensorReader` in the `ASPumpingSystem` view has no specified source port. The C&C model `PumpStation` satisfies these views.

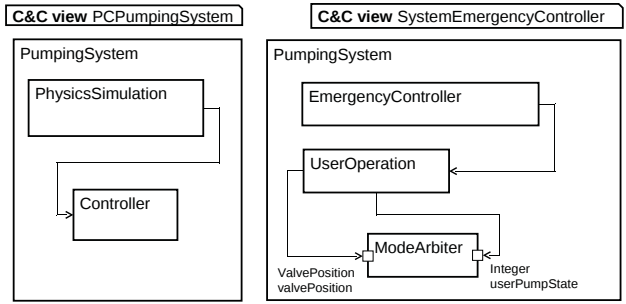

**Figure 3:** Two C&C views: `PCPumpingSystem` and `SystemEmergencyController`. The C&C model `PumpStation` does not satisfy these views.

connectors in the C&C model. In mathematical notation: `PumpStation` $\models$ `UserButton`.

Fig. 3 (left) shows a C&C view named `PCPumpingSystem`, which includes a connection from the `PhysicsSimulation` component to the `Controller` component, both within the `PumpingSystem` component. It is easy to see that the C&C model does not satisfy the view (we denote this by `PumpStation` $\not\models$ `PCPumpingSystem`). First, in the view, `PhysicsSimulation` is inside `PumpingSystem`, while in the C&C model it is not. Second, the connector from `PhysicsSimulation` to `Controller` shown in the view does not have a corresponding connector (or chain of connectors) in the C&C model.

Finally, Fig. 3 (right) shows a C&C view named `SystemEmergencyController`, which specifies structural properties related to a use case of emergency. Again, the pump station C&C model does not satisfy this view. First, the view includes the `EmergencyController` component, which does not exist in the C&C model. Second, the type `Integer` of the port named `userPumpState` of the `ModeArbiter` component, does not match the type `Boolean` of the port with the same name `userPumpState` in the C&C model. In mathematical notation: `PumpStation` $\not\models$ `SystemEmergencyController`.

Note that the views shown above crosscut the traditional boundaries of systems and sub-systems. They abstract the hierarchy (or parts of the hierarchy) away and instead focus on the components and connectors participating in a use case (e.g., the user pressing a button) or a certain concern (e.g., emergency). This crosscutting nature is a distinctive feature of C&C views.

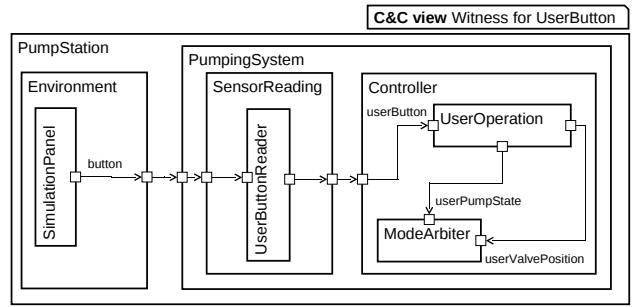

**Figure 4:** Generated witness for satisfaction of the `UserButton` view

We give the formal definitions for a C&C model, a view, and when does the former satisfy the latter, in Sect. 3.

## 2.2 Witness for Satisfaction/Non-Satisfaction

As mentioned above, given a C&C model and a view, we are interested in checking whether the former satisfies the latter. However, in practice, a Boolean answer is by far not enough. Rather, we look for concrete justifications for satisfaction or non-satisfaction. These should enable comprehension and in the case of non-satisfaction, point the engineer to the problems and assist her in correcting her design.

As an example, Fig. 4 shows a view which serves as a witness for showing that `PumpStation` $\models$ `UserButton`, that is, that the C&C model `PumpStation` satisfies the view `UserButton`. Note the complete hierarchy (excluding siblings) up until the least common parents of the components appearing in the view, and the chains of concrete connectors corresponding to the abstract connectors in the view, e.g., the chain from `SimluationPanel` to `ModeArbiter`.

As another example, Fig. 5 shows two views which serve as witnesses for showing that `PumpStation` $\not\models$ `PCPumpingSystem`, that is, that the C&C model `PumpStation` does not satisfy the view `PCPumpingSystem`. While in the view `PCPumpingSystem` the `PumpingSystem` component contains the `PhysicsSimulation` component, in the C&C model, as shown in the witness on the left, they are independent. While in `PCPumpingSystem` there is a connection from component `PhysicsSimulation` to component `Controller` (from unnamed port to unnamed port), in the C&C model, as shown in the witness on the right, `Controller` is not in the set of components reachable with a chain of connectors from `PhysicsSimulation`. Note that the two witnesses include annotations which explain, in natural language, the relevant reason for non-satisfaction.

In Sect. 4 we discuss the algorithms to decide satisfaction and generate informative witnesses like the ones we show above.

## 3. PRELIMINARIES

We briefly recall the structure of C&C models and views as defined in [24] (we give shortened definitions, for complete definitions see the technical report available from [3]).

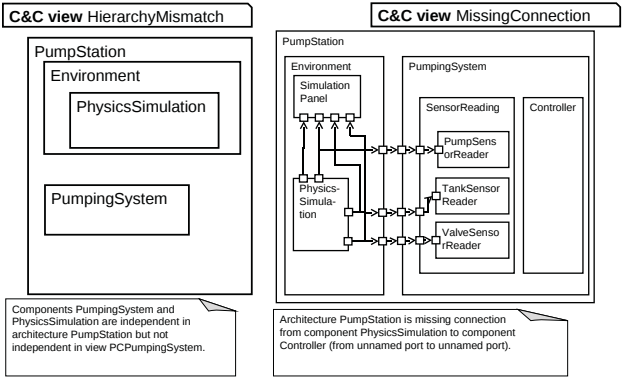

**Figure 5:** Two generated witnesses for non-satisfaction of the `PCPumpingSystem` view, including generated texts that explain the reason for non-satisfaction

### 3.1 Component and Connector Models

A C&C model is a structure
$cncm = \langle Cmps, Ports, Cons, Types, subs, ports, type\rangle$ where

- $Cmps$ is a set of named components, each of which has a set of ports $ports(cmp) \subseteq Ports$ and a (possibly empty) set of immediate subcomponents $subs(cmp) \subset Cmps$,
- $Ports$ is a disjoint union of input and output ports $Ports = PortsIn \cup PortsOut$ where each port $p \in Ports$ has a name, a type $type(p) \in Types$, and belongs to exactly one component $p \in ports(cmp)$,
- $Cons$ is a set of directed connectors, each of which connects two ports of the same type, which belong to two sibling components or to a parent component and one of its immediate subcomponents, and
- $Types$ is a finite set of type names.

Some additional well-formedness rules apply, e.g., that the subcomponents relation induced by $subs$ is a strict partial order, that every port has at most one incoming connector, and that port names are unique within their component. In addition, without loss of generality, we consider only C&C models with exactly one top component.

### 3.2 C&C Views

A C&C view is a structure $view = \langle Cmps, Ports, AbsCons, Types, subs, ports, type\rangle$ where

- $Cmps$ is a set of named components, each of which has a (possibly empty) set of ports $ports(cmp) \subseteq Ports$ and a (possibly empty) set of subcomponents $subs(cmp) \subset Cmps$,
- $Ports$ is a disjoint union of sets of input and output ports $Ports = PortsIn \cup PortsOut$ where each port $p \in Ports$ has a (possibly unknown) name, a (possibly unknown) type $type(p) \in Types \cup \bot$, and belongs to exactly one component $p \in ports(cmp)$,
- $AbsCons$ is a set of abstract connectors, each of which connects components (optionally) via ports of the same type or an unknown type, and
- $Types$ is a finite set of type names.

Note that in a C&C view, abstract connectors are not required to connect only two sibling components or a par-ent component and one of its immediate subcomponents. Again, the subcomponents relation is a strict partial order, but we do not restrict C&C views to have exactly one top component. We are now ready to define the semantics of our C&C view and C&C model, and specifically, when does the second satisfy the first.

A C&C model satisfies a C&C view iff the types, components, and ports mentioned in the second are contained in the first, the first respects the subcomponent relation induced by the second, two ports connected by an abstract connector in the second are connected by a chain of connectors in the first (respecting direction, names, and types), and all ports of a component in the second belong to the same component in the first with corresponding name, type and direction. More formally:

**Definition 1** ($cncm \models view$). A C&C model $cncm$ satisfies a C&C view $view$ iff:

- $view.Types \subseteq cncm.Types$, $view.Cmps \subseteq cncm.Cmps$, $view.Ports \subseteq cncm.Ports$,
- $\forall cmp_1, cmp_2 \in view.Cmps$: $cmp_1 \in view.subs(cmp_2)$ iff $cmp_1 \in cncm.subs^+(cmp_2)$ (we use $^+$ to denote the transitive closure),
- $\forall ac \in view.AbsCons \: \exists$ chain of connectors in $cncm$, $c_1, \ldots, c_n$ with $ac.srcCmp = c_1.srcCmp$ and $ac.tgtCmp = c_n.tgtCmp$ with matching port names and types, if specified, and
- $\forall cmp \in view.Cmps$:
  (1) $view.ports(cmp) \subseteq cncm.ports(cmp)$, and
  (2) $\forall p \in view.ports(cmp)$: $p \in view.PortsIn$ iff $p \in cncm.PortsIn \land view.type(p) \in \{\bot, cncm.type(p)\}$
  (similarly for unknown and given port names).

## 4. C&C VIEWS VERIFICATION

### 4.1 Problem Definition

The C&C views verification problem is defined as follows: given a C&C model $cncm$ and a C&C view $v$, decide whether $cncm \models v$. Moreover, in addition to the Boolean answer, we are interested in constructing minimal witnesses (we explain our notion of minimality below) that demonstrate it, as follows.

#### 4.1.1 A Witness for Satisfaction

In case the C&C model satisfies the view, the witness should be a minimal subset of the C&C model that is (1) by itself a well-formed C&C model, (2) contains all the components appearing in the view and their parent components up until their least common parent, (3) contains C&C model ports corresponding to all the ports appearing on all components in the view, and (4) contains C&C model connectors (and chains of connectors) representing all abstract connectors appearing in the view.

For example, recall the two views shown in Fig. 2. Both views are satisfied by the pump station C&C model. Fig. 4 shows the witness for satisfaction of the `UserButton` view (as discussed already in Sect. 2), and Fig. 6 shows the witness for satisfaction of the `ASPumpingSystem` view. Note the complete hierarchy (excluding siblings) up until the least common parents of the components appearing in the views, e.g., for `ASPumpingSystem`, the `SensorReading` component. Also note the chain of concrete connectors corresponding to the abstract connectors in the views, e.g., the chain from `PumpSensorReader` to `PumpActuator`.

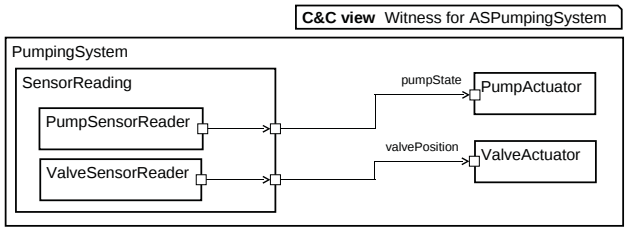

**Figure 6:** Witness for satisfaction of the `ASPumpingSystem` view

*4.1.2 Witnesses for Non-Satisfaction*

In case the C&C model does not satisfy the view, we are interested in a set of witnesses, each of which should explain one cause for non-satisfaction. Thus, the witnesses we are looking for are divided into four classes according to four reasons of non-satisfaction as follows:

- **Missing Component**: the view contains a component that does not exist in the C&C model;
- **Hierarchy Mismatch**: the view contains two components that in the C&C model are in a different containment relation (independent in view but not independent in the C&C model, not independent in the view but independent in the C&C model, not independent in both the view and the C&C model but in reverse containment relation);
- **Interface Mismatch**: the view contains a component with a port that does not exist in the C&C model (no full match of name, type, direction); and
- **Missing Connection**: the view contains an abstract connection that has no corresponding concrete chain of connectors in the C&C model.

**Remark 1.** Note that the classification of four reasons for non-satisfaction described above is complete, that is, given a C&C model *cncm* and a view *view*, if *cncm* $\not\models$ *view* then at least one of the reasons above holds. The correctness of our algorithm (see below) is based on this property. A formal proof is available in the technical report from [3].

Given the above classification of reasons, we require the generation of a *set of witnesses*, each of which is a minimal subset of the C&C model that is by itself a well-formed C&C model. The set should include the following witnesses:

- for each missing component, a witness that is an empty model annotated with the name of the missing component;
- for each hierarchy mismatch between two components, a witness that consists of the relevant pairs of components up to their least common parent in the C&C model but without siblings and without any connectors and ports;
- for each interface mismatch, a witness that consists of the relevant components in the C&C model (without subcomponents) with the relevant port (in case of type or direction mismatch) or their complete interface (in case no matching port was found) annotated with the port from the view that has no match; and

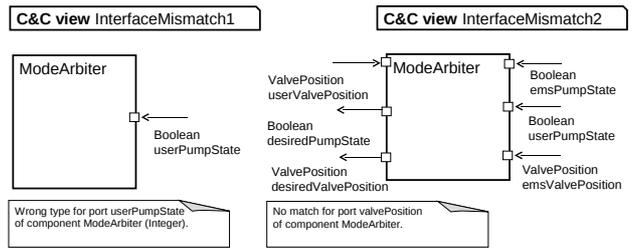

**Figure 7:** Non-satisfaction witnesses for the view `SystemEmergencyController`

- for each missing connection, a witness that consists of the relevant pairs of components up to their least common parent in the C&C model but without siblings and with all components reachable by connectors from the source component.

For example, as presented in Sect. 2, for the pump station C&C model and the view `PCPumpingSystem` (see Fig. 1 and Fig. 3 (left)), we consider the two witnesses shown in Fig. 5, corresponding to two reasons for non-satisfaction. First, the witness shown in Fig. 5 (left), asserting that `PumpingSystem` and `PhysicsSimulation` are independent, points to a hierarchy mismatch. Second, the witness shown in Fig. 5 (right), asserting that `Controller` is not in the set of components reachable with a chain of connectors from `PhysicsSimulation`, points to a missing connection.

As another example, for the pump station C&C model and the view `SystemEmergencyController` (see Fig. 1 and Fig. 3 (right)), we consider four witnesses, corresponding to four reasons for non-satisfaction, two interface mismatches, one missing component, and one missing connection.

**Remark 2.** Note that according to the above, each of the witnesses (for non-satisfaction as well as for satisfaction) is not only by itself a well-formed C&C model, but also, can be presented, technically, as a view (albeit with concrete containment and connectors) that is satisfied by the C&C model (indeed this can be checked by the same algorithm). This has two advantages. First, the engineer does not need to learn a new language in order to understand witnesses. Second, the same tools applied to C&C models and views, e.g., for presentation or further analysis, can be applied to witnesses.

To conclude, all witnesses have some common properties. First, each contains a single least common parent component, which is the top of the relevant sub-system (except for the case of a missing component). Second, each can be presented as a view with concrete containment and connectors. Finally, each witness includes complete port information of name, type, and direction (although the set of ports shown is not necessarily complete).

**Remark 3.** Interestingly, the witness for satisfaction of a witness (when considered as a view and checked against the same C&C model), is the witness itself. That is, a witness is itself the witness for its satisfaction. We consider this to be a nice idempotence property of our definition of views satisfaction and witnesses.

### 4.2 Algorithm Overview

We now give an overview of the algorithm we use. A detailed presentation of the algorithm, its correctness and com-plexity, and the templates mentioned below, are all available in a technical report from [3].

The input for the algorithm consists of a C&C model and a view. The output is a Boolean answer, whether the C&C model satisfies the view or not, and a set of one or more witnesses. The algorithm works by checking for reasons of non-satisfaction of the view and the C&C model in the four classes described in Sect. 4.1.2. The checks are done sequentially and independently; the answer that the C&C model satisfies the view is returned iff no reason for non-satisfaction is found.

Checking for the four non-satisfaction reasons is done using standard graph-traversing algorithms (depth-first search, breadth-first search, etc.) over the graphs induced by the inclusion relation between components and by the connectedness relation between ports in the model and the view. Checking for missing components and interface mismatches is straightforward. Checking for hierarchy mismatches is done by checking (1) for each component and each of its subcomponents in the view, whether the parent also contains the subcomponents in the C&C model, and (2) for each two components in the view that are not contained in one another, whether they are also independent in the model. Checking for missing connections is done by a breadth-first search of the connectors in the model.

The number of traversals is bounded by the total number of elements in the model and view at hand. Data collected during the traversals is used in the construction of the witnesses and in the instantiation of the natural-language templates for the textual explanations.

Witness construction is based on the data collected during the checking of the four reasons for non-satisfaction. In case of satisfaction, the witness is built by traversing the view and adding the corresponding elements from the model. In case of non-satisfaction, the input for the construction is the model and a single reason for non-satisfaction (e.g., a single pair of hierarchy mismatched components). Construction starts from the elements mentioned in the reason for non-satisfaction and adds required elements from the model.

### 4.2.1 Minimality

The generated witnesses for non-satisfaction are minimal in terms of number of components, connectors, and ports, because there is only one possible witness to construct in each case, by definition. For the witness of satisfaction, minimality is more subtle, because, as we explain below, there may be more than one possible correct witness.

To generate a small witness for $cncm \models view$, before adding a chain of connectors for an abstract connector or adding a port shown in the view to the witness, the algorithm checks whether the elements in the witness already provide a matching. If no match of the view's element exists in the witness, a match in the model is found and added to the witness. Note that this check is a heuristics to create small witnesses (when possible), in terms of the total number of concrete connectors and ports in the witnesses, but it does not guarantee that the generated witness is minimal to this measure. Also note that computing a shortest chain of connectors for each abstract connector does not guarantee a global minimum either, because in a minimal solution, some concrete connectors may potentially belong to multiple chains, i.e., be used to implement multiple abstract connectors.

Thus, computing minimal witnesses for satisfaction is possible but more complex and computationally expensive. In the current implementation we chose a fast computation based on the heuristics described above rather than a slower computation of a global minimum.

### 4.2.2 Generating Natural Language Descriptions

Finally, for each of the generated witnesses, in each of the four classes of non-satisfaction reasons, we generate a detailed description in natural language.

To generate natural language descriptions for the reasons for non-satisfaction we use plain-text, parametrized templates, one template for each non-satisfaction reason. The data collected by the algorithm (e.g., the name of a missing port) is used to assign values to the parameters in the template's instantiation.

For example, the template for hierarchy mismatch uses the following parameters: $cncm$, $view$, the hierarchy mismatch kind (three possible values, see Sect. 4.1.2), the parent component $cmp$, and the child component $subCmp$. For the second hierarchy mismatch kind, the template reads `"Components cmp.name and subCmp.name are independent in C&C model cncm.name but not independent in view view.name."`. Thus, for example, for the hierarchy mismatch between the pump station C&C model and the view `PCPumpingSystem`, as discussed above, the template's instantiation generates the following text: *"Components PumpingSystem and PhysicsSimulation are independent in C&C model PumpStation but not independent in view PCPumpingSystem"*.

The generated texts appear on the witnesses list in the Eclipse plug-in (see Sect. 5) and as a comment in the generated witness's document (see Fig. 5 and 7). They are intended to further help the engineer in identifying the reasons for non-satisfaction.

## 5. EVALUATION

The plug-in implementation and all specifications reported on below are available in supporting materials [3], together with screen captures and relevant documentation. All specifications can be inspected and all experiments can be reproduced. We encourage the interested reader to try it out.

Our evaluation includes tool support implementation, four example systems, performance and scalability experiments, and a user study, as described below.

### 5.1 Tool Support

We have implemented support for C&C views verification within a prototype Eclipse plug-in. The concrete syntax used is adapted from MontiArc [2].

Using the plug-in, the engineer can select two files, a C&C model and a view, and check whether the former satisfies the latter. In case of a positive answer, a witness for satisfaction is generated and presented as an annotated view in the main editing pane. In case of a negative answer, all witnesses for non-satisfaction are generated and listed in a hierarchical problems view titled *Witnesses for Non-Satisfaction*. The hierarchy in the problems view reflects the classification of non-satisfaction reasons described above, i.e., of missing component, hierarchy mismatch, interface mismatch, and missing connection. Clicking a specific entry in the problems view opens up the corresponding witness as an annotated view in the main editing pane. Fig. 8 shows a screen capture from the prototype plug-in.

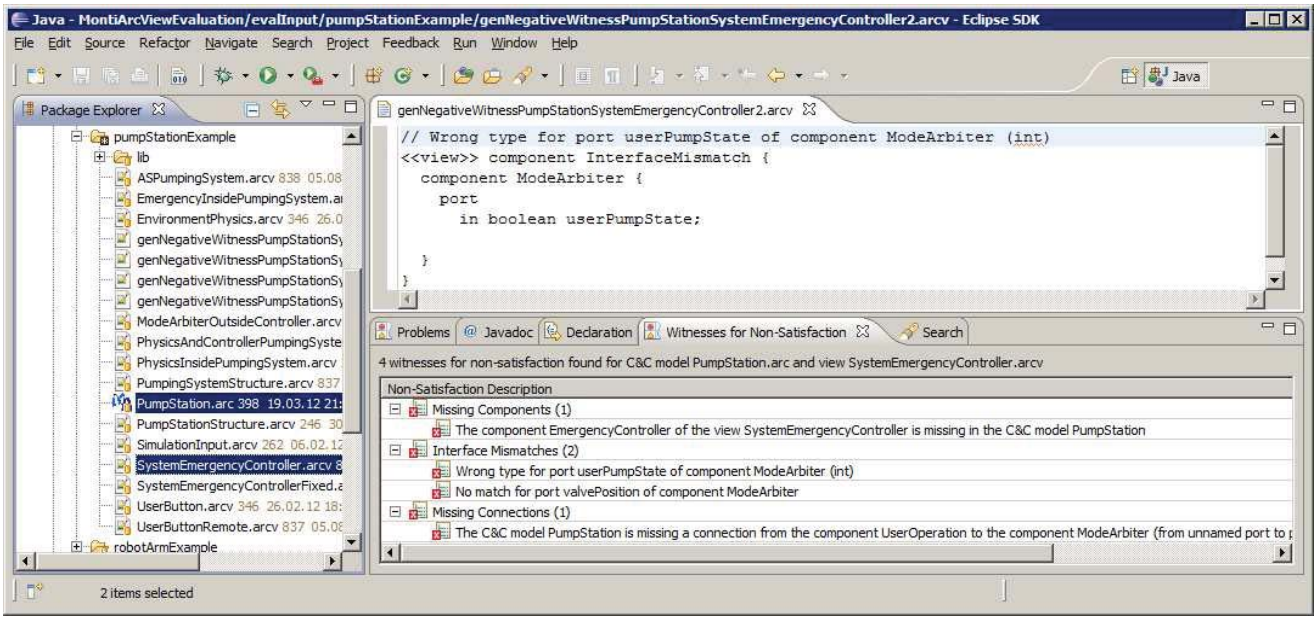

**Figure 8:** A screen capture from the prototype plug-in, after checking the C&C model PumpStation (shown in Fig. 1) against the view SystemEmergencyController (shown in Fig. 3, right). The lower pane shows the Eclipse problems view titled *Witnesses for Non-Satisfaction*, which provides a hierarchical list of the generated witnesses for non-satisfaction and their generated natural language descriptions. Four witnesses were generated, one for a missing component, two for interface mismatches, and one for a missing connection. The main editing pane shows one of the generated witnesses for interface mismatch.

## 5.2 Four Example Systems

We evaluated C&C views verification on four systems, taken from different sources and of different domains.

**Avionics System.** First, an AADL model of an avionics system, taken from [4] (specifically Avionics_System.aadl of the OSATE AADL Project). The avionics system C&C model is a high-level model of several avionics system subsystems. Since in this work we are only interested in the structure of the models, in our translation of this AADL architecture to a MontiArc architecture we have ignored the flows definitions but preserved the hierarchical structure and all ports and connectors. The avionics system C&C model has 6 components, 16 ports, and 8 connectors. The depth of the component hierarchy is 2.

Based on various use cases, related to the interaction between the avionics system's components, we created 9 views, satisfied and non-satisfied, with 1-6 components each. For example, one view gives an overview of the complete data flow in the system, declared using abstract connectors. This view does not provide additional information such as port names or types. Another view provides additional details about the communication between the Pilot_Display and its Page_Content_Manager, showing incoming and outgoing ports with their names and connectors. Verification produced up to 4 witnesses for each of the views we have defined.

**Bumper Bot.** Second, the software architecture of a Lego Mindstorms NXT [22] bumper bot. The bumper bot can power its left and right motors and detect obstacles in front of it (similar to the bumper car model from [27]). In addition, it is equipped with an emergency stop button. The bumper bot's objective is to go around obstacles and keep driving forward. As part of this example system, we have designed a set of views and a complete architecture for the bumper bot software in MontiArc. The bumper bot C&C model consists of 12 components, 28 ports, and 20 connectors. The depth of the component hierarchy is 3.

We defined 8 satisfied and non-satisfied views in total, each with 2-8 components. Of the 8 views, 5 are independent of the emergency stop feature and describe only the components and abstract connectors for fulfilling the main purpose of the bumper bot. The remaining 3 views exhibit components required by the emergency stop feature. For example, one of these adds a mode arbiter to the components and abstract connectors participating in the regular robot control. Another one shows the components and paths of signals used in case of an emergency stop, including the mode arbiter. Verification produced up to 2 witnesses for each of the views we have defined.

**Pump Station.** Third, the pump station architecture taken from an example system provided with the AutoFocus tool [1, 19] (the model we use as a running example in this paper). The physical pump station system consists of two water tanks connected by a pipeline system with a valve and a pump. The water level in the first water tank can rise (this is controlled by the environment). When the water level of the first tank rises to a critical level, the water has to be pumped to the the second water tank. The second water tank has a drain. The architecture presented in Fig. 1 also shows a model of the environment with a physics simulation, used to test the pumping system. The C&C model consists of 16 components, 67 ports, and 49 connectors. The depth of the component hierarchy is 4.

Based on several design decisions and relations we wanted to highlight and document, we created 10 views, each with 2-5 components. For example, one view gives an overview of the basic structure of the system and omits details about interfaces and connectors. Another view documents part of the connections between the actuators and their environment, hiding hierarchies and omitting elements not connected to the actuators. An additional view shows an undesired design where the simulation component is placed inside the pumping system. We have already described some of the views of this example system in Sect. 2. Verification produced up to 4 witnesses for each view.

**Robotic Arm.** Finally, we applied C&C views verification on a robotic arm architecture – specifically the rotational joint of a robotic arm, taken from an industrial system by VTT Tampere, Finland. The main components of the rotational joint's C&C model are a cylinder, a servo valve, a sensor, a joint limiter, and an actuator. The rotational joint is a subsystem of a robotic arm containing 8 rotational (identical copies) and translational joints in total. The robotic arm rotational joint C&C model consists of 8 components, 18 ports, and 16 connectors. The depth of the component hierarchy is 3.

Based on several requirements and partial knowledge or particular features, we created 11 C&C views, each with 1-5 components. Some views highlight the components necessary for the function of the joint while others document design alternatives on the placement of sensor and actuator components. Some of the views give an overview over related components with only few details of their interfaces or connectedness. Other views document complete interfaces of relevant components and some of their connections. Verification produced up to 4 witnesses for each view.

### 5.2.1 Lessons Learned from the Four Systems

In the four example systems, running times for checking satisfaction, including parsing and witness generation for satisfaction or non-satisfaction, were in all cases very fast. For example, running all the checks mentioned in all four systems above took less than 1 second, in total (on an ordinary laptop computer). This seems to be in line with the results of our synthetic experiments of performance and scalability (see below).

Moreover, running times for checking satisfaction of specifications was in all cases very fast too. Again, this comes at no surprise because checking satisfaction of a specification requires at most $k$ executions of our algorithm, where $k$ is the number of distinct views appearing in the specification.

The number of witnesses computed for all C&C views in all our four example systems was at most 4. We believe that this is due to the relatively small number of design decisions typically documented in a single view. The underspecification mechanisms of C&C views allow one to focus on only few elements per view to support comprehension. Another reason for the small number of witnesses is that by definition, all checks for non-satisfaction reasons – except for missing components – ignore components in the view that do not appear in the C&C model.

## 5.3 Performance and Scalability

To examine the performance and scalability of C&C views verification in handling large C&C models and views, we have experimented with synthesized models of different sizes and with related synthesized views where we have randomly applied various mutations. We describe the setup of our experiments below. The code to reproduce our experiments or further define and execute similar ones is available together with the rest of the evaluation materials in [3].

We have implemented a generator for random C&C models, which constructs random models based on the following parameters: number of components, maximal number of subcomponents per component, number of types of ports in the model, maximal number of ports and maximal number of connectors in the model.

We further implemented a generator for random views. For a given C&C model, the views generator works in two steps. First, it clones the C&C model and eliminates some components, ports, and connectors based on the following parameters: number of components to keep, maximal number of ports to keep, and maximal number of connectors to keep (the actual number of ports and connectors in the view depends also on the number of ports and connectors left on the components to keep). Moreover, chains of concrete connectors are replaced by corresponding abstract connectors. Second, it randomly applies one or more of the following mutations: changing the type of a port, renaming a component, renaming a port, and switching the names of two components. We have also randomly applied three benign mutations: eliminating the name of a port, eliminating the type of a port, and removing port information from the ends of a connector. We call these mutations benign because they preserve satisfaction. In all cases, we implemented the mutations in a way that guarantees that the resulting mutated view is syntactically valid (i.e., well-formed). For example, if a component is renamed, the new name is also applied to the connectors that end or start at this component (that is, the sets `Cmps` and `AbsCons` and the functions `subs` and `ports` are updated accordingly).

Below we show the results of executing C&C views verification, including the computation of non-satisfaction reasons and the generation of witnesses, on randomly generated models and mutated views of different sizes. Specifically, in the experiments we increased the number of components in the synthesized models from 20 to 200, the maximal number of direct subcomponents per component was fixed to 8, the maximal number of port types was fixed to 8, the total number of ports in the model was fixed to 8 times the number of components in the model, and the maximal number of connectors in the model was set to half of the number of ports. In one setup we set the number of components in the view to a fifth of the number of components in the model, and set the number of mutations to a third of the number of components in the view. In another setup we fixed the number of components in the view to 12 and the number of mutations to 6. In both setups, we set the number of ports and abstract connectors to keep to 2 times the number of components in the view. The first setup is designed to examine scalability. The second setup where the views are of fixed size is designed to be more realistic, based on our experience with using C&C models and views.

We executed all the experiments on an ordinary laptop (Dell Latitude E6500 laptop running Windows 7). We repeated the experiments 12 times for each C&C model size, from 20 to 200, for the two setups. Fig. 9 reports the average times (in milliseconds) to decide satisfaction and compute the reasons for non-satisfaction, for the two setups. Al-

102

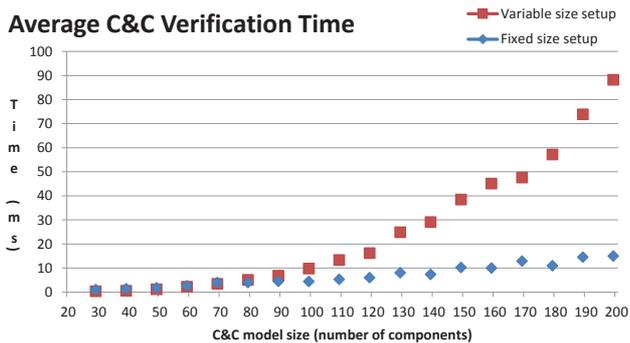

**Figure 9:** Average times (in milliseconds) to decide satisfaction and compute the reasons for non-satisfaction, for the two setups, the variable size setup where the view size is a fifth of the architecture size, and the fixed size setup where the view size is fixed to 12. Although the average times for the variable setup grow faster than the average times for the fixed setup, the absolute times recorded and the chart's growth clearly show that C&C views verification is feasible and scales well. See Sect. 5.3.

though the average times for the variable setup grow faster than the average times for the fixed setup, the absolute times recorded and the chart's growth clearly show that C&C views verification is feasible and scales well. Moreover, in our experiments, average and maximal times to generate a witness were 11 ms and 595 ms respectively in the first setup, and 5 ms and 768 ms respectively in the second setup.

The experiment results show that C&C views verification is feasible and scales well. Note that fast and scalable performance comes at no surprise, since our algorithms are polynomial in the size of the input C&C models and views.

## 5.4 User Study

We conducted a small user study to examine two high-level research questions: [RC1] is C&C verification difficult to do manually, and [RC2] are witnesses for satisfaction/non-satisfaction helpful. The study included a two-pages introduction on C&C views to read, 3 verification questions (each presenting one C&C view), and 3 questions about the usefulness of a set of witnesses that was presented to the user, all referring to a common C&C model. Two of the views in the first 3 questions had 2 non-satisfaction reasons each. The questions of each group were presented to the users in a random order to avoid a bias due to learning effect.

The study subjects were all CS graduate students or professional software engineers, all with some modeling background but no specific previous knowledge on our work on C&C views. No grades or other reward was involved. The study was anonymous and conducted online. We obtained complete set of answers from 17 subjects. The complete questionnaire and results are available from [3].

To answer [RC1] we measured the time spent on the first three questions, the correctness of the answers, and their completeness (identifying all reasons for non-satisfaction, where applicable). We also asked the subjects to report about their confidence in the correctness and completeness of their answers. The average (median) time to answer a verification question was 3.4 (2.9) minutes. 9 subjects (53%) identified non-satisfaction correctly and found all non-satisfaction reasons. 8 subjects (47%) missed at least one reason for non-satisfaction. Only 13 subjects (76%) identified satisfaction correctly (4 have 'identified' wrong non-satisfaction reasons). Only 3 subjects (18%) reported full confidence in the correctness of their answers and only 4 reported full confidence that they have identified all non-satisfaction reasons.

These results show that manual C&C verification is time-consuming and error prone. This justifies the need for automation.

To answer [RC2] we presented satisfaction results and witnesses (same C&C model as in the first three questions, but different views) to the subjects and asked them about the helpfulness of the witnesses. On average, 11.6 subjects (68%) reported to have found the witnesses we presented to them *helpful* or *very helpful* (top 2 out of 5 options). Only two subjects never found any of the witnesses helpful. 15 subjects found the witnesses helpful at least once.

These results are promising. Further investigation is required in order to identify which types of witnesses are more helpful than others, and how to improve witnesses helpfulness. Ideas for improvement we have received from the study subjects and other users include alternative witness constructions (e.g., not include the least common parent component), different witness presentation (e.g., using animation, or by visually overlaying the witness and the model to prevent the need for context switching), and richer textual explanations (e.g., adding text explicitly describing the elements shown in the witness).

### 5.4.1 Threats to Validity

Our set of 17 subjects is small and heterogeneous, e.g., in terms of previous modeling experience. The study involved a single small model (the pump station model) and only 3 verification questions and 3 witness usefulness questions based on views that we have designed. In the second part, we did not distinguish between different kinds of witnesses although their usefulness may vary. We also did not check for a correlation between the correctness of the answers in the first part to the perceived usefulness of witnesses in the second part. In the future we plan a larger study with more control on participants background and with more questions.

## 6. RELATED WORK

We discuss related work in the area of component and connector modeling and analysis, specifically related to abstraction mechanisms and verification. The key distinctive features of our work on C&C views verification are the focus on structure, the 'by example' and partial characteristics of the views, the expressive power and formal nature of the specification approach, and the generation of witnesses and textual explanations to justify the results of the automated analysis.

Many works investigate various analyses of C&C models (e.g., differencing and merging [5, 10, 20], composition [9], etc.), mainly in the context of architectures. None of these considers verification of structural properties expressed using views.

AADL [13] includes under-specification mechanisms similar to the ones available in C&C views. For example, AADL supports specifications with incomplete information of port types and with abstract flows, which show the source and sink of flows but not their complete path through the system.

Feiler et al. [13] explain that the motivation behind AADL's support for partial specifications is to allow some analysis (e.g., computing some metrics, checking syntactic correctness) already during early design, before all implementation details are known. We have not found any work on verifying the structure of AADL architectures against specifications (made of some kind of views or by other means).

Acme [14] allows one to specify first-order predicates on the structure of architectures, dealing, e.g., with connectedness between components. In AcmeStudio [6], these predicates are automatically evaluated against the architecture. The predicates are written as textual logical formulas, not as abstract models such as C&C views. Acme's predicate language is in some ways more expressive than C&C views, as it has the flexibility of first order logic and allows quantification over components, connectors, ports, and roles. However, the language neither supports the transitive subcomponent relation nor has constructs to crosscut the boundaries of the traditional implementation-based hierarchical decomposition of systems to subsystems, which is the very essence of C&C views. Finally, Acme verification provides only a Boolean answer while C&C views verification provides a set of witnesses. We have not seen similar witnesses for satisfaction or non-satisfaction of invariants in AcmeStudio.

More recently, Bhave et al. [8] have extended AcmeStudio to support structural consistency between heterogeneous models as architectural views, specifically for cyber-physical systems. View consistency is checked by verifying if a morphism exists between the two typed graphs. The work mentions reporting back to the user so she is able to "spot the inconsistent elements" as future work. In addition, the work discusses a single case study and provides no performance and scalability results. We currently do not deal with heterogeneous models, but instead focus on structural properties and on the abstraction of direct containment and connectivity. Unlike this work, we do report on several example systems, performance results, and a user study that examines the usefulness of our solution.

SysML [29, 30] is a general-purpose modeling language for systems engineering applications. The language is defined as an extension of a subset of the Unified Modeling Language (UML) [28], using UML's profiling mechanism. Previous work in our group [15, 16] described the use of views in the context of product lines, with a focus on the automotive domain, using SysML's internal block diagrams. SysML's internal block diagrams provide under-specification mechanisms, for example, to specify abstract connectors similar to the abstract connectors of C&C views. However, the question of verifying the structure of a C&C model against a view is discussed neither in these works nor in any other SysML related work we have found.

Behjati et al. [7] have extended SysML with AADL concepts, using a profile, to support comprehensive system architecture modeling and analysis. It may be possible to define C&C views within their extended language, so as to integrate it with standard, available tools.

Moriconi et al. [26] have investigated a notion of what they call refinement between architectural models, adding implementation details while preserving an isomorphic mapping (the implementation cannot add elements that are not mapped to elements in the architecture). A C&C model could be seen as a refinement of a C&C view that it satisfies. However, our notion of refinement would be fundamentally different as the C&C model may add elements (components, ports, connectors) that do not appear in the view.

Finally, from a broader, higher-level perspective, many behavioral specification languages exist and some of them have related verification methods and 'views'. For example, linear temporal logic (LTL) [23] is a behavioral specification language, and scenarios, expressed, e.g., using live sequence charts (LSC) [12, 18], may be considered as related behavioral views. As another example, in SysML, behavior is specified using activity diagrams and state machine diagrams.

In contrast, C&C views are intentionally limited to structural properties. In the behavioral case, a system's behavior is typically modeled using a state machine, and the behavioral properties this state machine needs to satisfy are expressed using LTL formulas or scenarios. Model-checking techniques can be used to check whether a state machine satisfies a behavioral property and provide a counterexample in case of a negative answer [11]. In the structural case, which we focus on in our present work, the structure of a system is described using a C&C model and the properties it needs to satisfy are expressed using structural views. The algorithm we present is used to check whether a C&C model satisfies a view. The witnesses we generate in case of non-satisfaction may be considered as the structural analogue to counterexamples in behavioral model checking; as in the behavioral case, the witnesses are meant to improve comprehension and to guide repairing; also, as in the behavioral case, where counterexamples may themselves be viewed as scenarios (at least in the context of LTL model checking), in the structural case we focus on here, witnesses for non-satisfaction are themselves C&C models.

An interesting and challenging possible direction for future work is to combine the structural and behavioral viewpoints into a single views language, with a related formal verification technique.

## 7. CONCLUSION

We presented an approach for the efficient verification of C&C models against C&C views. A unique feature of our work is the generation of small model witnesses and short natural language texts that formally justify and explain the verification results to the engineer. We reported on an evaluation of our work, both in terms of the performance and scalability of the algorithms, and in terms of its usefulness to engineers.

The work complements our previous work on synthesis from C&C views [24]. Several future research directions arise from our evaluation in Sect. 5, including investigating alternatives of witness construction, presentation, and richer textual explanations (see Sect. 5.4).

Our C&C models and views language focuses intentionally on basic, pure concepts. We believe this is good for the investigation of fundamental principles, like abstraction mechanisms for hierarchy and connectivity. To be applied in industrial contexts, these concepts should be integrated into concrete languages and application domains, e.g., as extensions of AADL and SysML, with first-class connectors that can be typed and refined, with a distinction between child-parent connectors and component-component connectors, with support for instantiation and quantification etc. We have started working with our industrial partners towards this direction.